\documentclass[lettersize,journal]{IEEEtran}
\usepackage{amsmath}
\usepackage{algorithmic}
\usepackage{algorithm}
\usepackage{array}
\usepackage[caption=false,font=normalsize,labelfont=sf,textfont=sf]{subfig}
\usepackage{textcomp}
\usepackage{stfloats}
\usepackage{url}
\usepackage{verbatim}
\usepackage{graphicx}
\usepackage{cite}
\usepackage{graphicx}
\usepackage{cite}
\usepackage{picinpar}
\usepackage{amsmath}
\usepackage{url}
\usepackage{flushend}
\usepackage[latin1]{inputenc}
\usepackage{colortbl}
\usepackage{soul}
\usepackage{multirow}
\usepackage{pifont}
\usepackage{color}
\usepackage{alltt}
\usepackage[hidelinks]{hyperref}
\usepackage{enumerate}
\usepackage{siunitx}
\usepackage{breakurl}
\usepackage{epstopdf}
\usepackage{pbox}
\usepackage{enumitem}
\usepackage{xcolor}
\usepackage{wrapfig}
\usepackage{amsfonts}
\usepackage{mathrsfs}

\newtheorem{assumption}{Assumption}
\newtheorem{lemma}{Lemma}
\newtheorem{remark}{Remark}

\begin{document}

\title{Distributed Fixed-Time Consensus Control for Multiple AUV Systems with Input Saturations}

\author{Mien Van, Yuzhu Sun, Stephen Mcllvanna, Minh-Nhat Nguyen, Federico Zocco,  Zhijie Liu, Hsueh-Cheng Wang
\thanks{
This work was partly supported by the Natural Environment Research Council,
United Kingdom [grant number NE/V008080/1] and by the Royal Society [grant number IEC/NSFC/211236 and grant number RGS/R1/221356]. (Corresponding author: Mien Van.)

Mien Van, Yuzhu Sun, Minh-Nhat Nguyen, Stephen Mcllvanna and Federico Zocco are with the School of Electronics, Electrical Engineering and Computer Science, Queen's University Belfast, Belfast, United Kingdom (e-mail: m.van@qub.ac.uk), Zhijie Liu is with School of Intelligence Science and Technology, University of Science and Technology Beijing, China (email: liuzhijie2012@gmail.com). Hsueh-Cheng Wang is with with the Department
of Electrical and Computer Engineering, National Chiao Tung
University (NCTU), Taiwan. }
\thanks{}}

\markboth{IEEE Transactions on Vehicular Technology}%
{Shell \MakeLowercase{\textit{et al.}}: A Sample Article Using IEEEtran.cls for IEEE Journals}


\maketitle

\begin{abstract}
This study proposes a new distributed control method based on an adaptive fuzzy control for multiple collaborative autonomous underwater vehicles (AUVs) to track a desired formation shape within a fixed time. First, a formation control protocol based on a fixed-time backstepping sliding mode control is designed, in which the consensus cooperative tracking errors for each AUV will be formulated. Then, to compensate for the saturated control torques, an adaptive auxiliary variable is introduced. Finally, a fixed-time adaptive fuzzy logic control (FLC) is derived to approximate the unknown dynamics, in which the adaptive laws of the FLC is derived such that the adaptive signals and errors can be convergent within a fixed time. The fixed time convergence is desired in practice because it provides an exciting property that the global convergence of the whole system is independent with the initial states of the AUVs. The computer simulation results for a consensus formation control of four AUVs show that the proposed formation control can provide high tracking performance with lower and smoother control efforts. 
\end{abstract}


\section{Introduction}
Robotics have been extensively applied for many challenging applications, ranging from manufacturing, agriculture, space and ocean applications \cite{robotM1}, \cite{robotM2}. In some applications, the use of single robots or autonomous vehicles has limited efficiency, due to the limitation of sensing, endurance and payload carrying. To increase the efficiency of robots and autonomous vehicles for these applications, a concept of multiple collaborative robotics or swam robotics have been introduced \cite{swarm}. For underwater environment, multiple collaborative autonomous underwater vehicles (AUVs) have shown their great efficiency for many challenging applications like seabed monitoring, wind turbine inspection, marine debris monitoring and cleaning, etc \cite{AUV1}. However, controlling multiple AUVs working collaboratively is not a trivial task because the effects of nonlinear dynamics, communication delay between AUVs, and the effects of underwater environmental disturbances, i.e., waves, currents, etc., become more severe in underwater environment \cite{AUV2}.  

Many elegant control methods have been investigated for increasing the tracking accuracy and robustness of multi-agent systems.  Optimal controllers using distributed optimization have been developed \cite{optimal1}. A safe optimal controller based on control barrier function (CBF) has been proposed in \cite{optimal2}. Another approach based on reinforcement learning has been developed for the collaborative control of multi-agent systems \cite{optimal3}.  Model predictive control (MPC) has been explored for multi-agent systems generally \cite{MPC1}, \cite{MPC2}, and for multiple AUV systems specifically \cite{MPCAUV}. Although optimal controllers and MPCs provide a good tracking performance when the full knowledge of the system can be known in advance, it is difficult to handle the disturbances and/or uncertainty components. In order to handle disturbance/uncertainty components, robust controllers have been extensively developed \cite{robust1}. Robust controllers are particularly efficient for the control system of AUVs due to their robustness against the nonlinear effects of underwater working conditions. Due to its strong robustness against disturbances (i.e., matched disturbances), sliding mode control (SMC) techniques have been developed \cite{SMC1}, \cite{SMC2}. Despite the advantages of high robustness, SMC generates high chattering, which can cause significant oscillations within the operation of mechanical systems. To reduce the chattering for SMC, a distributed bio-inspired SMC has been proposed for multiple AUVs \cite{SMC3}. However, the conventional SMCs do not provide finite time/fixed time convergence for the systems. To provide a finite time convergence, finite time consensus control methods have been developed for multi-agent systems \cite{finite1}, \cite{finite2}. To obtain both finite time convergence and higher robustness, finite time sliding mode controllers have also been introduced \cite{finiteSMC1}, \cite{finiteSMC2}. Finite time controllers have also been employed for single AUV system \cite{finiteM} and multiple AUV systems \cite{finiteAUV}, \cite{finiteAUV2}. In \cite{finiteAUV3}, a terminal SMC has been developed for the formation tracking control of multiple AUVs. The main drawback of the finite time controllers is that the convergence time of the system is dependent on the inital states of the systems. This issue, unfortunately, prevents the applicability of finite time controllers for many practical applications because, in practice, some initial states of some agents are unavailable or unknown. To overcome this drawback, fixed-time controllers have been studied recently \cite{fixedtime1}, \cite{fixedtime11}. The use of fixed-time controllers can provide a fixed-time convergence, which is independent with the initial states, for the multi-agent systems \cite{fixedtime2}, \cite{fixedtime3}, \cite{fixedtime4}. 

One of the issues that reduces the tracking performance of robotic systems and multi-agent systems is the effects of unknown components such as unknown system parameters, friction terms, and faults, etc \cite{fault1}. This becomes even more severe for AUVs due to the severe effects of external environmental disturbances, especially for multiple collaborative AUV systems \cite{learning}. To approximate the unknown components, many learning techniques have been extensively developed. An iterative learning method has been employed for multi-agent systems \cite{ilearning}. An adaptive NN has been developed for multi-agent systems \cite{NN1}, \cite{NN2}, and for multiple AUVs \cite{NNAUV}. Adaptive fuzzy logic controllers have also been developed to take the knowledge of human about the dynamic system into the design to increase the approximation performance of the FLC \cite{FLC1}, \cite{FLC2}. Adaptive fixed-time FLCs have been developed to preserve the advantages of both fixed-time convergence property and the approximation capacity of FLC in \cite{FxTFLC1}, \cite{FxTFLC2}. However, the adaptive laws of the existing fixed-time FLCs do not provide fixed-time convergence for the system. In practice, it is desired that all the adaptive laws of the system can be convergent within a fixed-time to guarantee the global convergence of the system within a fixed-time. This is the main motivation of this paper.

Input saturation is another important consideration in the design of practical controllers for single agent and multi-agent systems since, in practice, the control efforts of actuators (i.e., motors) are limited \cite{Sat1}, \cite{Sat2}. Many efforts have been spent to find an effective mechanism to mitigate the effects of input saturation. In general, to reduce the effects of saturated control torques, an auxiliary design system can be employed \cite{Sat3}, \cite{Sat4}. 
  
In summary, there are existing research gaps for formation tracking control for multiple AUVs, which will be addressed in this study: (i) the fixed-time convergence of the design of controllers for multiple AUV systems, (ii) the fixed-time convergence of the adaptive laws of the adaptive FLCs, (iii) the input saturation problem needs to be addressed within the design of distributed formation control of multiple AUV systems. To address the research gaps, a new fixed-time distributed formation tracking control for multiple AUV systems is proposed. The distributed fixed-time consensus formation will be derived based on a backstepping SMC method. To approximate the unknown components, an adaptive fixed-time FLC will be developed, in which  the adaptive laws of FLC will be derived such that it can be convergent within a fixed-time to guarantee a global fixed-time convergence for the system. Furthermore, an auxiliary adaptive function will be introduced into the fixed-time controller to compensate for the effects of the overhead control efforts. The effectiveness of the new control algorithm will be tested on a consensus formation of four AUVs  and compared with the counterpart distributed SMC based on a computer simulation. To highlight the novelties of this paper, we compare the proposed method with the existing approaches as follows:

\begin{itemize}
\item Unlike the existing distributed consensus formation controllers for multiple AUV systems \cite{SMC2}, this paper develops a fixed-time distributed formation algorithm for AUVs using a backstepping SMC method to preserve the merits of Lyapunov stability of the backstepping control, high robustness of SMC and bounded convergence time of the fixed-time control theory.  
\item Unlike the existing adaptive fixed-time fuzzy controllers \cite{FxTFLC1}, \cite{FxTFLC2}, which do not guarantee a fixed time convergence for the adaptive laws of FLC, this paper develops a new adaptive fixed time fuzzy law to guarantee the fixed time convergence of the adaptive weights of FLC. This ensures a global fixed time convergence of the whole collaborative multiple AUVs system.
\item Unlike the existing consensus formation controllers for multiple AUVs \cite{AUV1}, \cite{finiteAUV}, \cite{SMC2}, which do no consider the input saturation issues, this paper incorporates an adaptive auxiliary function into the fixed-time distributed consensus controller to handle the problem of saturated control efforts. 
\end{itemize}

\section{FIXED-TIME STABILITY and CONVERGENCE, FUZZY LOGIC, GRAPH THEORY AND PROBLEM FORMULATION}
\subsection {Fixed-time stability}
A typical nonlinear system can be represented as follows\cite{Pol2012}:
\begin{equation}
\begin{aligned}
\label {DD1}
\dot{\xi}(t)=f(\xi(t)), \quad \xi(t_0)=\xi_0, \quad \xi\in\Re^n
\end {aligned}
\end{equation}
where $f(\cdot):\Re^n\rightarrow\Re^n$ is a possibly discontinuous vector field. The fixed time convergence is determined for system (\ref{DD1}) when it  is globally finite-time stable and its convergent time is bounded regardless the initial states of the system, i.e., $\forall{\xi_0}\in\Re^n$,  $T(\xi_0)\leq{T_\text{max}}$ is satisfied, where ${T_\text{max}}$ is a positive constant.

\begin{lemma}[\cite{Pol2012}]
\label{lemma1} If a positive definite continuous function $V(\xi):\Re^n\rightarrow\Re$ for system  (\ref{DD1}) satisfies $\dot{V}(\xi)\leq{-\chi_1V^\varrho(\xi)-\chi_2V^\varsigma(\xi)}$ for some $\chi_1>0$, $\chi_2>0$, $\varrho>1$, and $0<\varsigma<1$, then system (\ref{DD1}) is determined as a globally fixed-time stable system. The convergence time can be calculated independently with the initial states of system  (\ref{DD1}) as follows:
\begin{equation}
\begin{aligned}
\label {D2}
T(\xi_0)\leq\frac{1}{\chi_1(\varrho-1)}+\frac{1}{\chi_2(1-\varsigma)}.
\end {aligned}
\end{equation}
\end{lemma}

\begin{lemma}[\cite{Pol2012}]
\label{lemma2}
If a positive definite continuous function $V(\xi):\Re^n\rightarrow\Re$ for system  (\ref{DD1}) satisfies $\dot{V}(\xi)\leq{-\chi_1V^p(\xi)-\chi_2V^q(\xi)}+\varphi$ for some $\chi_1>0$, $\chi_2>0$, $p>1$, $0<q<1$, and $0<\varphi<\infty$, then system (\ref{DD1}) is called a practically fixed-time stable system. Futhermore, the solution of system (\ref{DD1}) has a residual set:
\begin{equation}
\begin{aligned}
\label {D3}
\lim_{t\to{T}}\xi| \Vert{\xi}\Vert \leq{\min}\{\chi_1^{\frac{-1}{p}}\left(\frac{\varphi}{1-\kappa}\right)^{\frac{1}{p}}, \chi_2^{\frac{-1}{q}}\left(\frac{\varphi}{1-\kappa}\right)^{\frac{1}{q}}\}
\end {aligned}
\end{equation}
where $\kappa$ satisfies $0<\kappa<1$. The settling time can be calculated independenly with the initial states of the system as follows:
\begin{equation}
\begin{aligned}
\label {D4}
T(\xi_0)\leq\frac{1}{\chi_1\kappa(1-p)}+\frac{1}{\chi_2\kappa(q-1)}.
\end {aligned}
\end{equation}
\end{lemma}

\subsection{Fuzzy Logic System}
Given a vector of input, i.e., $Z=(z_1,z_2,...,z_n)^T\in\Re^n$ and an output variable, i.e., ${y}=f({Z})\in\Re$, a fuzzy logic system can be used to map from the input to the output. The fuzzy rules of fuzzy logic system can be described as:
 \begin{equation}
\label {fuz1}
Rule \: \textit{j}: \text{If}\: z_1\: \text{is}\: A_1^j\: \text{and}\: ...\: \text{and}\:z_n\: \text{is}\: A_n^j\: \text{then}\: y\: \text{is}\: B^j
\end{equation}
 where $A_1^j$, $A_2^j$,..., $A_n^j$ and $B^j$ represent fuzzy sets. The fuzzy output can be obtained as:
 \begin{equation}
\label {fuz2}
y={}^{\sum\limits_{j=1}^{h}{{{w}_{j}}}\left( \prod\limits_{i=1}^{n}{{{\mu }_{A_{i}^{j}}}({{z}_{i}})} \right)}/{}_{\sum\limits_{j=1}^{h}{\left( \prod\limits_{i=1}^{n}{{{\mu }_{A_{i}^{j}}}({{z}_{i}})} \right)}}={{\text{w}}^{T}}\text{ }\!\!\Psi\!\!\text{ }({Z})
\end{equation}
 where $h$ specifies the number of fuzzy rules used, and ${{\mu }_{A_{i}^{j}}}({{z}_{i}})$ represents the membership function of ${z}_{i}$. $\text{w}={{\left[ {{w}_{1}},{{w}_{2}},..,{{w}_{h}} \right]}^{T}}$  represents the fuzzy weights, and $\text{ }\!\!\Psi\!\!\text{ }({Z})={{\left[ {{\Psi }_{1}}({Z}),{{\Psi }_{2}}({Z}),\dots,{{\Psi }_{h}}({Z}) \right]}^{T}}$  is a fuzzy basis vector, where its elements ${{\Psi }_{j}}({Z})$  can be described as

\begin{equation}
\label {fuz3}
{{\Psi }_{j}}({Z})={}^{\prod\limits_{i=1}^{n}{{{\mu }_{A_{i}^{j}}}({{z}_{i}})}}/{}_{\sum\limits_{j=1}^{h}{\left( \prod\limits_{i=1}^{n}{{{\mu }_{A_{i}^{j}}}({{z}_{i}})} \right)}}.
\end{equation}
 
\begin{lemma}[\cite{FxTFLC1, FxTFLC2}]
Let ${f(Z)}$ be a continuous function on a compact set $\Omega \in {{\Re }^{n}}$, there exists a fuzzy logic system, i.e., ${{\text{w}}^{T}}\text{ }\!\!\Phi\!\!\text{ }({Z})$, such that
\begin{equation}
\label {fuz4}
\underset{{Z}\in \Omega }{\mathop{\sup }}\,\left| {f(Z)}-{{\text{w}}^{T}}\text{ }\!\!\Psi\!\!\text{ }({Z}) \right|\le \bar{\varrho}
\end{equation}
where $\bar{\varrho}$ is the fuzzy minimum approximation error, $\text{ }\!\!\Psi\!\!{ (Z)}={{{\left[ {{\Psi }_{1}}({Z}),{{\Psi }_{2}}({Z}),...,{{\Psi }_{h}}({Z}) \right]}^{T}}}/{\sum\limits_{j=1}^{h}{{{\Psi }_{j}}({Z})}}$  is the fuzzy basis function vector.
\end{lemma}

\subsection {Graph theory}
Consier a directed graph $G=\{\Lambda,\Xi\}$ is used to describe the formation shape among a group of AUV vehicles, where $\Lambda=\{\nu_1, \nu_2,...,\nu_N\}$ denotes $N$ AUV followers, $\Xi\subseteq{\Lambda\times{\Lambda}}$ is the set of edges. $A=[\alpha_{ij}]\in\Re^{N\times{N}}$ denotes the weight of the edges, where $\alpha_{ij}=\alpha_{ji}>0$ if there is an edge between AUVs $i$ and $j$, i.e., $(\nu_j,\nu_i)\in{\Xi}$, and $\alpha_{ij}=\alpha_{ji}=0$ otherwise. Let $B=\text{diag}\{b_1,...,b_N\}$, where $b_i>0$ indicates that the follower AUV $i$ can receive the direct command signals from the AUV leader; under other conditions $b_i=0$. The Laplacian matrix $L=[l_{i,j}]\in\Re^{N\times{N}}$ with $l_{i,j}=-\alpha_{i,j}$ for $i\neq j$, and $l_{i,i}=\sum_{j=1}^N{\alpha_{i,j}}$. It is assumed that the graph $G$ is undirected and connected, and the desired trajectory information from the virtual leader will be transfered to at least one AUV, and thus not all the elements of B indentify to zero. Therefore, $L+B>0$.

\subsection{Dynamics of AUVs and Control Objective}
In this paper, a control method that can form the operations of $N$ AUVs with the dynamics described in (\ref{D1}) in a consensus manner will be derived.
\begin{equation}
\begin{aligned}
\label {D1}
&\dot{\eta}_i=J_i(\eta_{2,i})\upsilon_i,\\
&M_i\dot{\upsilon}_i+C_i(\upsilon_i)\upsilon_i+D_i(\upsilon_i)\upsilon_i=u_i\left(\tau_i(t)\right)+d_i(t,\eta_i,\upsilon_i)
\end{aligned}
\end{equation}
where $\eta_i=[\eta_{1,i},\eta_{2,i}]^T\in\Re^{6\times{1}}$, $\eta_{1,i}=[x_i, y_i, z_i]^T\in\Re^{3\times{1}}$, $\eta_{2,i}=[\phi_i, \theta_i, \psi_i]^T\in\Re^{3\times{1}}$ denote the position and orientation of $i$-th AUV, respectively. $\upsilon_i=[\upsilon_{1,i}, \upsilon_{2,i}]^T\in\Re^{6\times1}$, $\upsilon_{1,i}=[\upsilon_{x,i}, \upsilon_{y,i}, \upsilon_{z,i}]^T\in\Re^{3\times1}$, $\upsilon_{2,i}=[\omega_{x,i}, \omega_{y,i}, \omega_{z,i}]^T\in\Re^{3\times1}$ represents the translational and rotational velocities of $i$-th AUV, respectively. $u_i\left(\tau_i(t)\right)\in\Re^{6\times1}$, which will be described in (\ref{B15}), represents the control effort subject to saturation nonlinearity for the $i$-th AUV. The description of the inertia matrix $M_i\in\Re^{6\times6}$, the Coriolis and centripetal matrix $C_i(\upsilon_i)\in\Re^{6\times6}$, the hydrodynamic matrix $D_i(\upsilon_i)\in\Re^{6\times6}$ and the Jacobian matrix $J_i(\eta_{2,i})$ can be found in \cite{SMC3}. $d_i(t,\eta_i,\upsilon_i)\in\Re^{6\times1}$ denotes the lumped model uncertainty and disturbance component in the system.

\textit{Control Objective:} The objective of a distributed consensus controller is to design an appropriate controller for each AUV with the dynamics (\ref{D1}) so that the group of AUVs can: (i) form a desired formation shape, and (ii) follow a predefined trajectory, which is known as a virtual leader, within a fixed time. The desired formation shape of a group of AUVs can be determined by a specific relative postures, i.e., position and orientation, between AUVs.

\section{Fixed-time Backstepping Sliding Mode Control Design for Consensus Formation Tracking Control}

Let $\eta_i^d$, $\dot{\eta}_i^d$ and $\ddot{\eta}_i^d$ be the desired position, velocity and acceleration of the virtual leader. Define the position and orientation tracking errors between the objective trajectories and the reference trajectory for $i$-th AUV $(i\in\Gamma, \Gamma=\{1,...,N\})$ as follows:
\begin{equation}
\begin{aligned}
\label {2}
&\varepsilon_{1,i}=\sum_{j\in{\Gamma}}^{} \alpha_{ij} (\eta_i-\eta_j-\delta_{ij})+b_i(\eta_i-\eta_i^d-\delta_{id})\\
&\dot{\varepsilon}_{1,i}=\sum_{j\in{\Gamma}}^{} \alpha_{ij} (\dot{\eta}_i-\dot{\eta}_j)+b_i(\dot{\eta}_i-\dot{\eta}_i^d).
\end {aligned}
\end{equation}
Here, $\alpha_{ij}\geq{0}$ and $b_i\geq{0}$ are defined as in section II.C. $\delta_{ij}$ indicates the relative position and orientation between $i$-th AUV and $j$-th AUV ($j\in{\Gamma}$). $\delta_{id}$ denotes the relative posture between the $i$-th AUV and the reference trajectory (i.e., the virtual leader). All the AUVs are expected to have the same velocity and acceleration as the desired reference trajectory.

Differentiating the velocity of tracking error $\dot{\varepsilon}_{1,i}$ with respect to time, we have:
\begin{equation}
\begin{aligned}
\label {3}
\ddot{\varepsilon}_{1,i}=\sum_{j\in{\Gamma}}^{} \alpha_{ij} (\ddot{\eta}_i-\ddot{\eta}_j)+b_i(\ddot{\eta}_i-\ddot{\eta}_i^d)
\end {aligned}
\end{equation}
where $\ddot{\eta}_i$ and $\ddot{\eta}_j$ represent the acceleration of $i$-th AUV and its neighbors $j\in{\Gamma}$, respectively. Based on (\ref{D1}), the dynamic model of the $i$-th AUV can be expressed as:
\begin{equation}
\begin{aligned}
\label {4}
\ddot{\eta}_i=&\Phi_i\left(\upsilon_i,\eta_{i}\right)\upsilon_i+J_i\left(\eta_{2,i}\right)\Pi_iu\left(\tau_i(t)\right)\\
&+J_i(\eta_{2,i})\Pi_id_i(t,\eta_i,\upsilon_i)
\end {aligned}
\end{equation}
where,\\
$\Pi_i=M_i^{-1}$ and $\Phi_i(\upsilon_i,\eta_{i})=\dot{J}_i(\eta_{2,i})-J_i(\eta_{2,i})\Pi_iC_i(\upsilon_i)-J_i(\eta_{2,i})\Pi_iD_i(\upsilon_i)$.
\\
For facilitating the design of controllers later, the following matrices are defined:\\
$\bar{\Phi}(\upsilon,\eta)=\text{diag}\{\Phi_1(\upsilon_1, \eta_1), ...,\Phi_N(\upsilon_N, \eta_N)\}$,\\
$\bar{\Pi}=\text{diag}\{\Pi_1,...,\Pi_N\}$,\\
$\bar{J}(\eta_2)=\text{diag}\{J_1(\eta_{2,1}),...,J_N(\eta_{2,N})\}$,\\
$u\left(\tau(t)\right)=[u_1\left(\tau_1(t)\right), u_2\left(\tau_2(t)\right),...,u_N\left(\tau_N(t)\right)]^T$,\\
$d=[d_1(t,\eta_1,\upsilon_1),d_2(t,\eta_2,\upsilon_2),...,d_N(t,\eta_N,\upsilon_N)]^T$.\\

Therefore,
\begin{equation}
\begin{aligned}
\label {4_2}
\ddot{\eta}=\bar{\Phi}(\upsilon,\eta)\upsilon+\bar{J}(\eta_2)\bar{\Pi}u\left(\tau(t)\right)+\bar{J}(\eta_2)\bar{\Pi}d.
\end {aligned}
\end{equation}

\begin{assumption}
The disturbance term $J_i(\eta_{2,i})\Pi_id_i(t,\eta_i,\upsilon_i)$ is bounded by the positive constant  $\tilde{\lambda}_i$:
\begin{equation}
\begin{aligned}
\label {5}
\|J_i(\eta_{2,i})\Pi_id_i(t,\eta_i,\upsilon_i)\|\leq\tilde{\lambda}_i,{}{}{}i\in\Gamma.
\end {aligned}
\end{equation}
The parameter  $\tilde{\lambda}_i$ typically depends on the internal model uncertainties and external environmental disturbances (i.e., marine environment) of the vehicles.
\end{assumption}

Letting the following variables:
\begin{equation}
\begin{aligned}
\label {6}
\bar{\varepsilon}_1=[\varepsilon_{1,1}, \varepsilon_{1,2},...,\varepsilon_{1,N}]^T,\\
\bar{\varepsilon}_2=[\dot{\varepsilon}_{1,1}, \dot{\varepsilon}_{1,2},...,\dot{\varepsilon}_{1,N}]^T,
\end {aligned}
\end{equation}
and\\
\begin{equation}
\begin{aligned}
\label {7}
\ddot{\eta}=[\ddot{\eta}_1, \ddot{\eta}_2,...,\ddot{\eta}_N]^T.
\end {aligned}
\end{equation}

Adding the results in (\ref{3}), (\ref{6}) and (\ref{7}) to form the overall error dynamics as:
\begin{equation}
\begin{aligned}
\label {8}
&\dot{\bar{\varepsilon}}_1=\bar{\varepsilon}_2,\\
&\dot{\bar{\varepsilon}}_2=\left(L+B\right)\left(\ddot{\eta}-\textbf{1}_N\otimes\ddot{\eta}^d\right),\\
\end {aligned}
\end{equation}
where $\otimes$ denotes the Kronecker product between two matrices. $\textbf{1}_N$ stands for an $N\times{1}$ vector with unitary elements.

Then, based on (\ref{8}), a fixed time backstepping SMC can be designed as follows:
\\
\\
\textbf{Step 1}: The first sliding surface is selected as:
\begin{equation}
\begin{aligned}
\label {B2}
s_1(t)=\bar{\varepsilon}_1(t).
\end {aligned}
\end{equation}

Differentiating (\ref{B2}) yields
\begin{equation}
\begin{aligned}
\label {B3}
\dot{s}_1(t)=\alpha_s(t),
\end {aligned}
\end{equation}
where $\alpha_s(t)=\dot{\bar{\varepsilon}}_1(t)$ is identified as the virtual control of the system (\ref{B3}).

To stabilise the sliding surface $s_1(t)$, the following virtual control input is designed:
\begin{equation}
\begin{aligned}
\label {B4}
\alpha_s=-\left(k_1s_1+k_2s_1^{\gamma}+k_3s_1^{\iota}\right),
\end {aligned}
\end{equation}
where $k_1>0$, $k_2>$ and $k_3>0$ , and $0<\gamma<1$ and $\iota>1$.

Consider a candidate Lyapunov function below:
\begin{equation}
\begin{aligned}
\label {B5}
V_1&=\frac{1}{2}s_1^T{s}_1.\\
\end {aligned}
\end{equation}

Adding the result in (\ref{B4}) into the derivative of (\ref{B5}), we obtain:
\begin{equation}
\begin{aligned}
\label {B6}
\dot{V}_1&=s_1^T\dot{s}_1\\
&=-s_1^T\left(k_1s_1+k_2s_1^{\gamma}+k_3s_1^{\iota}\right)\\
&=-k_1s_1^Ts_1-k_2\left(s_1^Ts_1\right)^{\frac{\gamma+1}{2}}-k_3\left(s_1^Ts_1\right)^{\frac{\iota+1}{2}}\\
&\leq{-k_2\left(s_1^Ts_1\right)^{\frac{\gamma+1}{2}}-k_3\left(s_1^Ts_1\right)^{\frac{\iota+1}{2}}}\\
&\leq-2^\frac{\gamma+1}{2}k_2\left(\frac{1}{2}s_1^Ts_1\right)^\frac{\gamma+1}{2}-2^\frac{\iota+1}{2}k_3\left(\frac{1}{2}s_1^Ts_1\right)^\frac{\iota+1}{2}.
\end {aligned}
\end{equation}
\\
\textbf{Step 2}: Define the second sliding surface:
\begin{equation}
\begin{aligned}
\label {B7}
s_2=\bar{\varepsilon}_2-(L+B)\alpha_s.
\end {aligned}
\end{equation}

The derivative of $s_2$ is:
\begin{equation}
\begin{aligned}
\label {B8}
\dot{s}_2&=\dot{\bar{\varepsilon}}_2-\left(L+B\right)\dot{\alpha}_s\\
&=\left(L+B\right)\left(\ddot{\eta}-\textbf{1}_N\otimes\ddot{\eta}^d-\dot{\alpha}_s\right).\\
\end {aligned}
\end{equation}

A candidate Lyapunov function is selected as
\begin{equation}
\begin{aligned}
\label {B9}
V_2=\frac{1}{2}s_2^T{s}_2.\\
\end {aligned}
\end{equation}

Adding the result in (\ref{B8}) into the derivative of (\ref{B9}), we obtain:
\begin{equation}
\begin{aligned}
\label {B10}
\dot{V}_2&=s_2^T\dot{s}_2\\
&=s_2^T\left(\left(L+B\right)\left(\ddot{\eta}-\textbf{1}_N\otimes\ddot{\eta}^d-\dot{\alpha}_s\right)\right)\\
&=s_2^T\left((L+B)\left(\bar{\Phi}(\upsilon,\eta)\upsilon+\bar{J}(\eta_2)\bar{\Pi}u(\tau(t)) \right. \right.\\
&\left. \left.+\bar{J}(\eta_2)\bar{\Pi}d-\textbf{1}_N\otimes\ddot{\eta}^d-\dot{\alpha}_s\right)\right).\\
\end {aligned}
\end{equation}

Based on (\ref{B10}), the backstepping sliding mode controller can be taken as
\begin{equation}
\begin{aligned}
\label {B11}
u(\tau(t))&=[\bar{J}(\eta_2)\bar{\Pi}]^{-1}\left(-\bar{\Phi}(\upsilon,\eta)\upsilon+\textbf{1}_N\otimes\ddot{\eta}^d+\dot{\alpha}_s+\tau^{'}\right),
\end {aligned}
\end{equation}
where,
\begin{equation}
\begin{aligned}
\label {B11_2}
\tau^{'}&=-\beta_s\text{sign}(s_2)-k_8s_2-k_9s_2^\gamma-k_{10}s_2^\iota,
\end {aligned}
\end{equation}
where $k_8, k_9, k_{10}$ are positive constants. $\beta_s$ is chosen to be $\beta_s>\tilde{\lambda}$, where $\tilde{\lambda}=[\tilde{\lambda}_1, \ldots, \tilde{\lambda}_N]^T$, which were defined as in Assumption 1.

Inserting the control input in (\ref{B11}) and (\ref{B11_2}) into (\ref{B10}), we have:
\begin{equation}
\begin{aligned}
\label {B12}
\dot{V}_2&=s_2^T\dot{s}_2\\
&=s_2^T\left(L+B\right)\left({-\beta_s\text{sign}(s_2)-k_8s_2-k_9s_2^{\gamma+1}} \right.\\
&\left. {-k_{10}s_2^{\iota+1}+\bar{J}(\eta_2)\bar{\Pi}d} \right)\\
&\leq-2^\frac{\gamma+1}{2}\left(L+B\right)k_9\left(\frac{1}{2}s_2^Ts_2\right)^\frac{\gamma+1}{2}\\
&-2^\frac{\iota+1}{2}\left(L+B\right)k_{10}\left(\frac{1}{2}s_2^Ts_2\right)^\frac{\iota+1}{2}.
\end {aligned}
\end{equation}

\textbf{Step 3}: We define a compounded candidate Lyapunov function:
\begin{equation}
\begin{aligned}
\label {B13_1}
{V}=V_1+V_2.
\end {aligned}
\end{equation}

Differentiating (\ref{B13_1}) yields:
\begin{equation}
\begin{aligned}
\label {B13}
\dot{V}&=\dot{V}_1+\dot{V}_2\\
&\leq-2^\frac{\gamma+1}{2}k_2\left(\frac{1}{2}s_1^Ts_1\right)^\frac{\gamma+1}{2}-2^\frac{\iota+1}{2}k_3\left(\frac{1}{2}s_1^Ts_1\right)^\frac{\iota+1}{2}\\
&-2^\frac{\gamma+1}{2}(L+B)k_9\left(\frac{1}{2}s_2^Ts_2\right)^\frac{\gamma+1}{2}\\
&-2^\frac{\iota+1}{2}(L+B)k_{10}\left(\frac{1}{2}s_2^Ts_2\right)^\frac{\iota+1}{2}\\
&\leq{-2^{\frac{\gamma+1}{2}}}\zeta_1\left(V_1^{\frac{\gamma+1}{2}}+V_2^{\frac{\gamma+1}{2}}\right)\\
&-{2}^{\frac{\iota+1}{2}}\zeta_2\left(V_1^{\frac{\iota+1}{2}}+V_2^{\frac{\iota+1}{2}}\right)\\
&\leq{-2^{\frac{\gamma+1}{2}}}\zeta_1\left(V^{\frac{\gamma+1}{2}}\right)-{2}^{\frac{\iota+1}{2}}\zeta_2\left(V^{\frac{\iota+1}{2}}\right).
\end {aligned}
\end{equation}
where $\zeta_1=\text{min}\left(k_2,\left(L+B\right)k_9\right)$ and $\zeta_2=\text{min}\left(k_3,\left(L+B\right)k_{10}\right)$.

Therefore, thanks to Lemma \ref{lemma1}, the global fixed-time convergence can be established for the system (\ref{B13}), and the reaching time can be obtained as:
\begin{equation}
\begin{aligned}
\label {B14}
T\leq{\frac{2}{{\zeta_12^{\frac{\gamma+1}{2}}}(1-\gamma)}}+\frac{2}{{2}^{\frac{\iota+1}{2}}\zeta_2(\iota-1)}.
\end {aligned}
\end{equation}

\section{Distributed Backstepping Fuzzy Sliding Mode Controller with Input Saturation}
The backstepping SMC presented in section III has two main shortcomings: (i) the bigger sliding gain is chosen based on Assumption 1, for which if the disturbance is big, the controller provides a big chattering in the system, (ii) the saturated control torque effects have not been considered. In this section, we introduce an auxiliary variable and a fuzzy approximation to overcome these shortcomings. 

The designed control input $\tau_i(t)\in\Re^n$ is affected by the saturation nonlinearity and can be expressed as \cite{Sat2}:
\begin{equation}
\begin{aligned}
\label {B15}
u_i(\tau_i(t))=\begin{cases}
    \text{sign}(\tau_i(t))\tau_{\text{max}_i}, & |\tau_i(t)|\geq{\tau_{\text{max}_i}},\\
    \tau_i(t), & |\tau_i(t)|<{\tau_{\text{max}_i}},
  \end{cases}
\end {aligned}
\end{equation}
where ${\tau_{\text{max}_i}}$ represents the maximum control torque allowed for joint $i$.

Furthermore, considering the input saturation, the saturated control torque can be approximated by
\begin{equation}
\begin{aligned}
\label {B16}
u_i(\tau_i)=g_i(\tau_i)+\varsigma_i(\tau_i),
\end {aligned}
\end{equation}
where $g_i(\tau_i)$ is a smooth function. $\varsigma_i(\tau_i)$ is the bounded approximation error. $g_i(\tau_i)$ can be chosen as \cite{Sat2}:
\begin{equation}
\begin{aligned}
\label {B17}
g_i(\tau_i)&=\tau_{\text{max}_i}\times{\text{tanh}}\left(\frac{\tau_i}{\tau_{\text{max}_i}}\right)\\
&=\tau_{\text{max}_i}\frac{e^{\tau_i/\tau_{\text{max}_i}}-e^{-\tau_i/\tau_{\text{max}_i}}}{e^{\tau_i/\tau_{\text{max}_i}}+e^{-\tau_i/\tau_{\text{max}_i}}}.
\end {aligned}
\end{equation}

The approximation error $\varsigma_i(\tau_i)$ is bounded by
\begin{equation}
\begin{aligned}
\label {B18}
|\varsigma_i(\tau_i)|=|u_i(\tau_i)-g_i(\tau_i)|\leq\tau_{{\text{max}_i}}(1-\text{tanh}(1))=\bar{\Delta}_i.
\end {aligned}
\end{equation}
Let: $g(\tau(t))=[g_1(\tau_1(t)), g_2(\tau_2(t)),...,g_N(\tau_N(t))]^T$, $\varsigma(\tau(t))=[\varsigma_1(\tau_1(t)), \varsigma_2(\tau_2(t)),...,\varsigma_N(\tau_N(t))]^T$. To compensate for the saturated controller, an adaptive auxiliary variable $\mu$ is introduced as:
\begin{equation}
\begin{aligned}
\label {B19}
\dot{\mu}=-\mu+\bar{J}(\eta_2)\bar{\Pi}\left(g(\tau)-\tau\right).
\end {aligned}
\end{equation}
For this controller, \textbf{Step 1} is as in section III. \textbf{Step 2} will be re-designed as follows:\\
\\
\textbf{Step 2}: The error variable $s_2$ can be redefined as
\begin{equation}
\begin{aligned}
\label {B20}
s_2=\bar{\varepsilon}_2-(L+B)\alpha_s-(L+B)\mu.
\end {aligned}
\end{equation}
The derivative of $s_2$ can be computed as
\begin{equation}
\begin{aligned}
\label {B21}
\dot{s}_2&=\dot{\bar{\varepsilon}}_2-(L+B)\dot{\alpha}_s-(L+B)\dot{\mu}\\
&=\left(L+B\right)\left(\ddot{\eta}-\textbf{1}_N\otimes\ddot{\eta}^d-\dot{\alpha}_s-\dot{\mu}\right)\\
&=\left(L+B\right)\left(\bar{\Phi}(\upsilon,\eta)\upsilon+\bar{J}(\eta_2)\bar{\Pi}u(\tau) \right.\\
&\left. +\bar{J}(\eta_2)\bar{\Pi}d-\textbf{1}_N\otimes\ddot{\eta}^d-\dot{\alpha}_s-\dot{\mu}\right)\\
&=(L+B)\left(\bar{\Phi}(\upsilon,\eta)\upsilon+\bar{J}(\eta_2)\bar{\Pi}u(\tau)\right.\\
&\left. +\bar{J}(\eta_2)\bar{\Pi}d-\textbf{1}_N\otimes\ddot{\eta}^d-\dot{\alpha}_s\right.\\
&\left.+\mu-\bar{J}(\eta_2)\bar{\Pi}\left(g(\tau)-\tau\right)\right)\\
&=(L+B)\left(\bar{\Phi}(\upsilon,\eta)\upsilon+\bar{J}(\eta_2)\bar{\Pi}\left(\varsigma(\tau\right)+d) \right.\\
&\left. -\textbf{1}_N\otimes\ddot{\eta}^d+\mu+\bar{J}(\eta_2)\bar{\Pi}\tau-\dot{\alpha}_s\right).
\end {aligned}
\end{equation}

By using a FLC to approximate the lumped uncertainty and disturbance $\bar{J}(\eta_2)\bar{\Pi}(\varsigma(\tau)+d)$, the derivative of the error $s_2$ in (\ref{B21}) can be represented as
\begin{equation}
\begin{aligned}
\label {45}
\dot{s}_2=&(L+B)\left(\bar{\Phi}(\upsilon,\eta)\upsilon+W^{*T}\Psi(Z)+\epsilon(t) \right.\\
&\left.-\textbf{1}_N\otimes\ddot{\eta}^d+\mu+\bar{J}(\eta_2)\bar{\Pi}\tau-\dot{\alpha}_s \right),
\end {aligned}
\end{equation}
where $\epsilon(t)=\bar{J}(\eta_2)\bar{\Pi}\left(\varsigma(\tau)+d(t,\eta,\upsilon)\right)-W^{*T}\Phi(Z)$. From (\ref{5}) and (\ref{B18}) and Lemma 3, we can obtain $||\epsilon(t)||\leq\bar{\epsilon}$, where $\bar{\epsilon}>0$.

A candidate Lyapunov function is defined as
\begin{equation}
\begin{aligned}
\label {46}
V_3(s_2,\tilde{\theta})=\frac{1}{2}s_2^Ts_2+\frac{1}{2}\tilde{\theta}^T\tilde{\theta},
\end {aligned}
\end{equation}
where $\tilde{\theta}= \theta-\hat{\theta}$ is the weight approximation error with $\theta=\max_{h\in{H}}\Vert{W^*}\Vert$, and $\hat{\theta}$ is the approximation of $\theta$.

The derivative of $V_3$ in (\ref{46}) can be computed as:
\begin{equation}
\begin{aligned}
\label {47}
\dot{V}_3&=s_2^T\dot{s}_2-\tilde{\theta}^T\dot{\hat{\theta}}\\
&=s_2^T(L+B)\left(\bar{\Phi}(\upsilon,\eta)\upsilon+W^{*T}\Psi(Z)+\epsilon(t) \right.\\
&\left. -\textbf{1}_N\otimes\ddot{\eta}^d+\mu+\bar{J}(\eta_2)\bar{\Pi}\tau-\dot{\alpha}_s\right)-\tilde{\theta}^T\dot{\hat{\theta}}.
\end {aligned}
\end{equation}

Let $F_{sum}=\bar{\Phi}(\upsilon,\eta)\upsilon-\textbf{1}_N\otimes\ddot{\eta}^d+\mu-\dot{\alpha}_s$. Applying inequality principle, we have
\begin{equation}
\begin{aligned}
\label {48}
s_2^TW^{*T}\Psi(Z)\leq{\frac{1}{2}+\frac{1}{2}s_2^2\theta\Psi(Z)^T\Psi(Z)}.
\end {aligned}
\end{equation}

Then, $\dot{V}_3$ becomes:
\begin{equation}
\begin{aligned}
\label {49}
\dot{V}_3&\leq{s_2}^T(L+B)\left({\bar{J}(\eta_2)\bar{\Pi}\tau+F_{sum}+\epsilon(t)}\right.\\
&\left.{+\frac{1}{2}s_2^2\hat{\theta}\Psi(Z)^T\Psi(Z)}\right)+\frac{1}{2}(L+B)-\tilde{\theta}^T\dot{\hat{\theta}}\\
&\leq{s_2^T}(L+B)\left({\bar{J}(\eta_2)\bar{\Pi}\tau+F_{sum}+\epsilon(t)}\right.\\
& \left.{+\frac{1}{2}s_2^2\hat{\theta}\Psi(Z)^T\Psi(Z)}\right)+\frac{1}{2}(L+B)\\
&+\tilde{\theta}^T\left(\frac{1}{2}(L+B)s_2^2\Psi(Z)^T\Psi(Z)-\dot{\hat{\theta}}\right).\\
\end {aligned}
\end{equation}

Based on (\ref{49}), a distributed control input is designed as:
\begin{equation}
\begin{aligned}
\label {491}
\tau&=\left(\bar{J}\left(\eta_2\right)\bar{\Pi}\right)^{-1}\left(-F_{sum}-\frac{1}{2}s_2^2\hat{\theta}\Psi(Z)^T\Psi(Z)\right.\\
&\left. -\beta_s{\text{sign}}(s_2)-k_8s_2-k_9s_2^\gamma-k_{10}s_2^\iota \right),
\end {aligned}
\end{equation}
where $\beta_s$ is selected such that $\beta_s>\bar{\epsilon}$.

The adaptive law of the FLC can be selected as
\begin{equation}
\begin{aligned}
\label {50}
\dot{\hat{\theta}}=\frac{1}{2}\left(L+B\right)s_2^2\Psi(Z)^T\Psi(Z)-w_1\hat{\theta}^{\gamma}-w_2\hat{\theta}^{\iota}.
\end {aligned}
\end{equation}

Therefore,
\begin{equation}
\begin{aligned}
\label {51}
\dot{V}_3&\leq{s_2^T(L+B)\left(-k_8s_2-k_9s_2^\gamma-k_{10}s_2^\iota\right)}\\
&+\frac{1}{2}(L+B)+\tilde{\theta}^T\left(w_1\hat{\theta}^{\gamma}+w_2\hat{\theta}^{\iota}\right).
\end {aligned}
\end{equation}

Using inequality:
\begin{equation}
\begin{aligned}
\label {52}
\tilde{\theta}\hat{\theta}^{\gamma}\leq{l_1\theta^{1+\gamma}-l_2\tilde{\theta}^{1+\gamma}},
\end {aligned}
\end{equation}

\begin{equation}
\begin{aligned}
\label {52b}
\tilde{\theta}\hat{\theta}^{\iota}\leq{l_1\theta^{1+\iota}-l_2\tilde{\theta}^{1+\iota}}.
\end {aligned}
\end{equation}

Taking the results in (\ref{51}), (\ref{52}) and (\ref{52b}) together, we have:
\begin{equation}
\begin{aligned}
\label {53}
\dot{V}_3&\leq\left(L+B\right)\left(-k_8s_2-k_9s_2^{\gamma+1}-k_{10}s_2^{\iota+1}\right)\\
&+\frac{1}{2}(L+B)+w_1\left({l_1\theta^{1+\gamma}-l_2\tilde{\theta}^{1+\gamma}}\right)\\
&+w_2\left({l_1\theta^{1+\iota}-l_2\tilde{\theta}^{1+\iota}}\right)\\
&\leq(L+B)\left({-k_8s_2-k_9s_2^{\gamma+1}-k_{10}s_2^{\iota+1}}\right)\\
&-l_2\left(w_1\tilde{\theta}^{1+\gamma}+w_2\tilde{\theta}^{1+\iota}\right)+\sigma,
\end {aligned}
\end{equation}
where:
\begin{equation}
\begin{aligned}
\label {54}
\sigma=\frac{1}{2}\left(L+B\right)+l_1\left(w_1{\theta}^{1+\gamma}+w_2{\theta}^{1+\iota}\right).
\end {aligned}
\end{equation}

Therefore, 
\begin{equation}
\begin{aligned}
\label {55}
\dot{V}_3&\leq{-k_9\left(L+B\right)s_2^{\gamma+1}}-l_2w_1\tilde{\theta}^{1+\gamma}\\
&-k_{10}\left(L+B\right)s_2^{\iota+1}-l_2w_2\tilde{\theta}^{1+\iota}+\sigma\\
&\leq{-2^{\frac{\gamma+1}{2}}}\nu_1{V_3}^{\frac{1+\gamma}{2}}{-2^{\frac{\iota+1}{2}}}\nu_2{V_3}^{\frac{1+\iota}{2}}+\sigma.\\
\end {aligned}
\end{equation}

Here,
\begin{equation}
\begin{aligned}
\label {56}
\nu_1=\text{min}\left(k_9\left(L+B\right),l_2w_1\right)
\end {aligned}
\end{equation}
\begin{equation}
\begin{aligned}
\label {57}
\nu_2=\text{min}\left(k_{10}\left(L+B\right),l_2w_2\right)
\end {aligned}
\end{equation}
Thus, according to Lemma \ref{lemma2}, the value $s_2$ and $\tilde{\theta}$ will converge to zero. The convergence time can be calculated as $T\leq\frac{2}{\nu_1{(1-\gamma)}}+\frac{2}{\nu_2{(\iota-1)}}$.
\\
\\
\textbf{Step 3}: Define a candidate Lyapunov function:
\begin{equation}
\begin{aligned}
\label {58}
{V}=V_1+V_3.
\end {aligned}
\end{equation}

The derivative of the above Lyapunov function is
\begin{equation}
\begin{aligned}
\label {59}
\dot{V}&\leq{-2^{\frac{\gamma+1}{2}}}\lambda_{min}\{k_2,k_5\}V_1^{\frac{\gamma+1}{2}}{-2^{\frac{\iota+1}{2}}}\lambda_{min}\{k_3,k_6\}V_1^{\frac{\iota+1}{2}}\\
&{-2^{\frac{\gamma+1}{2}}}\nu_1V_3^{\frac{\gamma+1}{2}}{-2^{\frac{\iota+1}{2}}}\nu_2V_3^{\frac{\iota+1}{2}}+\sigma\\
&\leq{-2^{\frac{\gamma+1}{2}}}\chi_1\left(V_1^{\frac{\gamma+1}{2}}+V_3^{\frac{\gamma+1}{2}}\right)\\
&-{2}^{\frac{\iota+1}{2}}\chi_2\left(V_1^{\frac{\iota+1}{2}}+V_3^{\frac{\iota+1}{2}}\right)+\sigma\\
&\leq{-2^{\frac{\gamma+1}{2}}}\chi_1\left(V^{\frac{\gamma+1}{2}}\right)-{2}^{\frac{\iota+1}{2}}\chi_2\left(V^{\frac{\iota+1}{2}}\right)+\sigma,\\
\end {aligned}
\end{equation}
where $\chi_1=\text{min}\{\nu_1, \lambda_{min}\{k_2,k_5\}\}$ and $\chi_2=\text{min}\{\lambda_{min}\{k_3,k_6\}, \nu_2\}$.

Therefore, according to Lemma {\ref{lemma2}, the global fixed-time convergence of the system is  guaranteed. The settling time can be calculated as:
\begin{equation}
\begin{aligned}
\label {60}
T\leq{\frac{2}{{\chi_12^{\frac{\gamma+1}{2}}}\kappa(1-\gamma)}}+\frac{2}{{\chi_2}2^{\frac{\iota+1}{2}}\kappa(\iota-1)}.
\end {aligned}
\end{equation}

\begin{remark}
The employment of $sign$ function in (\ref{491}) generates a chattering in the system. In order to reduce the chattering, the controller (\ref{491}) can be revised as
\begin{equation}
\begin{aligned}
\label {492}
\tau&=\left(\bar{J}(\eta_2)\bar{\Pi}\right)^{-1}\left(-F_{sum}-\frac{1}{2}s_2^2\hat{\theta}\Psi(Z)^T\Psi(Z) \right.\\
&\left. -\beta_s(\frac{s_2}{||s_2||+\epsilon_1})-k_8s_2-k_9s_2^\gamma-k_{10}s_2^\iota \right),
\end {aligned}
\end{equation}
\end{remark}
where $\epsilon_1$ is a small positive number.

\section{Results and Discussions}
In this section, we validate the performance of the proposed algorithm. The dynamic model of each vehicle is described as in (\ref{D1}), where the parameters are selected as in Table \ref{table_1} \cite{SMC3}.

\begin{table}
\caption{Parameters used in the simulation of $i$th AUV ($i\in\{1,2,3,4\}$)}
	\centering
	\label{table_1}
\begin{tabular}{l l l l}
\hline\hline\\[-0.5ex]
Parameters &$Value$&Parameters &$Value$\\[0.5ex] \hline\\[-0.5ex]

$m_i$& $20$&$I_{x,i}$ & $20$\\

$I_{y,i}$ & $30$&$I_{z,i}$ & $35$\\

$\iota_{\upsilon{x},i}$&$-8$&$\iota_{\upsilon{y},i}$&$-10$\\
 $\iota_{\upsilon{z},i}$&$-9$&$\iota_{\dot{\upsilon}{x},i}$&$-7$\\
$\iota_{\dot{\upsilon}{y},i}$&$-8$&$\iota_{\dot{\upsilon}{z},i}$&$-6$\\
$\iota_{{\omega}{x},i}$&$-0.2$&$\iota_{{\omega}{y},i}$&$-0.25$\\
$\iota_{{\omega}{z},i}$&$-0.15$&$\iota_{\dot{\omega}{x},i}$&$-20$\\
$\iota_{\dot{\omega}{y},i}$&$-30$&$\iota_{\dot{\omega}{z},i}$&$-35$\\
\hline\hline 
\end{tabular}
\end{table}

Fig. \ref{fig.1} illustrates the connection between AUVs and the virtual leader, and $\alpha_{12}=a_{21}=\alpha_{23}=\alpha_{32}=\alpha_{34}=\alpha_{43}=1$. As illustrated in Fig. \ref{fig.1}, in this considered communication topology, the desired trajectory will be communicated and given to the AUV-1, i.e., $b_1=1$. Therefore, the $L$ and $B$ matrices can be calculated as:
\begin{equation}
\begin{aligned}
\label {61}
L=\begin{bmatrix}
1 &-1 & 0 & 0\\
-1 & 2 & -1 &0\\
0  & -1& 2&- 1\\
0 & 0 & -1 &1
\end{bmatrix},
B=\begin{bmatrix}
1 &0 & 0 & 0\\
0 & 0 & 0 &0\\
0  & 0& 0&0\\
0 & 0 & 0 &0
\end{bmatrix}.
\end {aligned}
\end{equation}

The moving trajectory of the virtual leader is selected as $\eta^d(t)=[30-30e^{-t}, 5t, 2t, 0, 0, 0]^T$. The desired posture between AUVs are given by $\delta_{12}=[0, 10, 0]^T$,   $\delta_{21}=[0, -10, 0]^T$,  $\delta_{23}=[-10, 0, 0]^T$,  $\delta_{32}=[10, 0, 0]^T$,  $\delta_{34}=[0, -10, 0]^T$, and  $\delta_{43}=[0, 10, 0]^T$. All the vehicles have the same orientation. The relative distance between the virtual leader and AUV 1 is $\delta_{1d}=[20, 0, 0]^T$. It is assumed that the four AUVs will start from the initial positions: $\eta_1(0)=[2, 3, 3, 0.3, 0, 0.2]^T$, $\eta_2(0)=[2.5, 3.5, 3, 0.2, 0, 0.25]^T$, $\eta_3(0)=[2, 3, 3, 0.3, 0, 0.2]^T$, $\eta_4(0)=[3, 3, 2, 0.3, 0, 0.2]^T$, and $\upsilon_i=0_{6\times{1}}, i\in\{1, 2, 3, 4\}$ is set for the initial velocities of AUVs.
 
The disturbance term is assumed to be: 
\begin{equation}
\begin{aligned}
\label {63}
 d_i(t,\eta_i,\upsilon_i)=&[2.5\sin(t)-0.5\upsilon_{xi}^2-0.7\sin(\upsilon_{xi}\upsilon_{yi}),\\
&2.5\cos(t)+0.1\upsilon_{xi}^2+0.5\sin(\upsilon_{yi}), \\
&2.5\sin(t)+0.7\upsilon_{xi}^2+0.8\sin(\upsilon_{zi}),\\
 &0.5\sin(t)+0.2\upsilon_{\phi{i}}^3,\\
 &0.5\cos(t)-0.2\upsilon_{\theta{i}}^2,\\
 &0.5\sin(t)-0.4\upsilon_{\Phi{i}}^3,]^T,\\
(i\in\{1, 2, 3, 4\}).
\end {aligned}
\end{equation}
Note that the above parameters are selected to be quite similar to the parameters used in \cite{SMC3} to facilite the comparison later. However, in this experiment, the disturbance term (\ref{63}) is modeled to be more severe to include both environmental disturbances and model uncertainties. 

The selected parameters, which were chosen based on a trial-and-error procedure, for the proposed controller in this simulation are $k_1=k_8=5$, $k_2=k_3=k_9=k_{10}=0.4$, $w_1=w_2=1$, $\beta_s=20$. The parameters $\gamma=5/7$ and $\iota=7/5$. The control efforts are saturared by $\tau_\text{max}=300 Nm$. 
The proposed controller uses the below membership functions, which were tuned based on a trial-and-error procedure:\\

$
{{\mu }_{A_{i}^{1}}}=\exp \left( {-{{\left( {{Z}_{i}}+7 \right)}^{2}}}/{4}\; \right), {{\mu }_{A_{i}^{2}}}=\exp \left( {-{{\left( {{Z}_{i}}+5 \right)}^{2}}}/{4}\; \right), 
{{\mu }_{A_{i}^{3}}}=\exp \left( {-{{\left( {{Z}_{i}}+3 \right)}^{2}}}/{4}\; \right),  {{\mu }_{A_{i}^{4}}}=\exp \left( {-{{\left( {{Z}_{i}}+1 \right)}^{2}}}/{4}\; \right),
{{\mu }_{A_{i}^{5}}}=\exp \left( {-{{\left( {{Z}_{i}}+0 \right)}^{2}}}/{4}\; \right),  {{\mu }_{A_{i}^{6}}}=\exp \left( {-{{\left( {{Z}_{i}}-1 \right)}^{2}}}/{4}\; \right),
{{\mu }_{A_{i}^{7}}}=\exp \left( {-{{\left( {{Z}_{i}}-3 \right)}^{2}}}/{4}\; \right),  {{\mu }_{A_{i}^{8}}}=\exp \left( {-{{\left( {{Z}_{i}}-5 \right)}^{2}}}/{4}\; \right),
{{\mu }_{A_{i}^{9}}}=\exp \left( {-{{\left( {{Z}_{i}}-7 \right)}^{2}}}/{4}\; \right).$\\

The input of the FLC is $Z_i=[\eta_i, \upsilon_i]^T$. To reduce chattering, the controller (\ref{492}) is used and $\epsilon_1=0.01$.

In order to highlight the superior performance of the proposed controller, it is analysed in a comparison with the distributed SMC \cite{SMC3}. The SMC can be designed as in Appendix A. The sliding gain of the SMC is selected as $\beta_0=200$. Note that the SMC \cite{SMC3} has not considered the effects of the input saturation in the design. The tracking performances of the proposed controller are shown in Figs. \ref{fig.2}, \ref{fig.3}, \ref{fig.4}, \ref{fig.5}, while the performances of the SMC are shown in Figs. \ref{fig.6}, \ref{fig.7}, \ref{fig.8}, \ref{fig.9}. In particular, Fig. \ref{fig.2} shows the formation shape of the four AUVs under the proposed controller. Compared with the formation shape of the AUVs under the SMC controller shown in Fig. \ref{fig.6}, the proposed controller provides faster and smoother convergence, as shown in Fig. \ref{fig.2}. Figs. \ref{fig.3} and \ref{fig.4} show the tracking errors of $\varepsilon_{1,i}, (i=1,2,3,4)$ and $\varepsilon_{2,i}, (i=1, 2, 3, 4)$ under the proposed controller, which are convergent to zero. The errors $\varepsilon_{1,i}$ and $\varepsilon_{2,i}$ under the effects of the SMC are shown in Figs. \ref{fig.7} and \ref{fig.8}, respectively. Comparing  Figs. \ref{fig.7} and \ref{fig.8} and Figs. \ref{fig.3} and \ref{fig.4}, respectively, show that the SMC provides faster convergence for the tracking errors. However, this is because the sliding gain of the SMC was chosen to be a bigger value. Consequently, it leads to a much higher, possibly physically unrealisable, control efforts, as shown in Fig. \ref{fig.9}. Fig. 5 shows the control efforts of the proposed controller, which are smoother and bounded by the $\tau_\text{max}$.

\begin{figure}[!t]\centering
	\includegraphics[width=8.5cm]{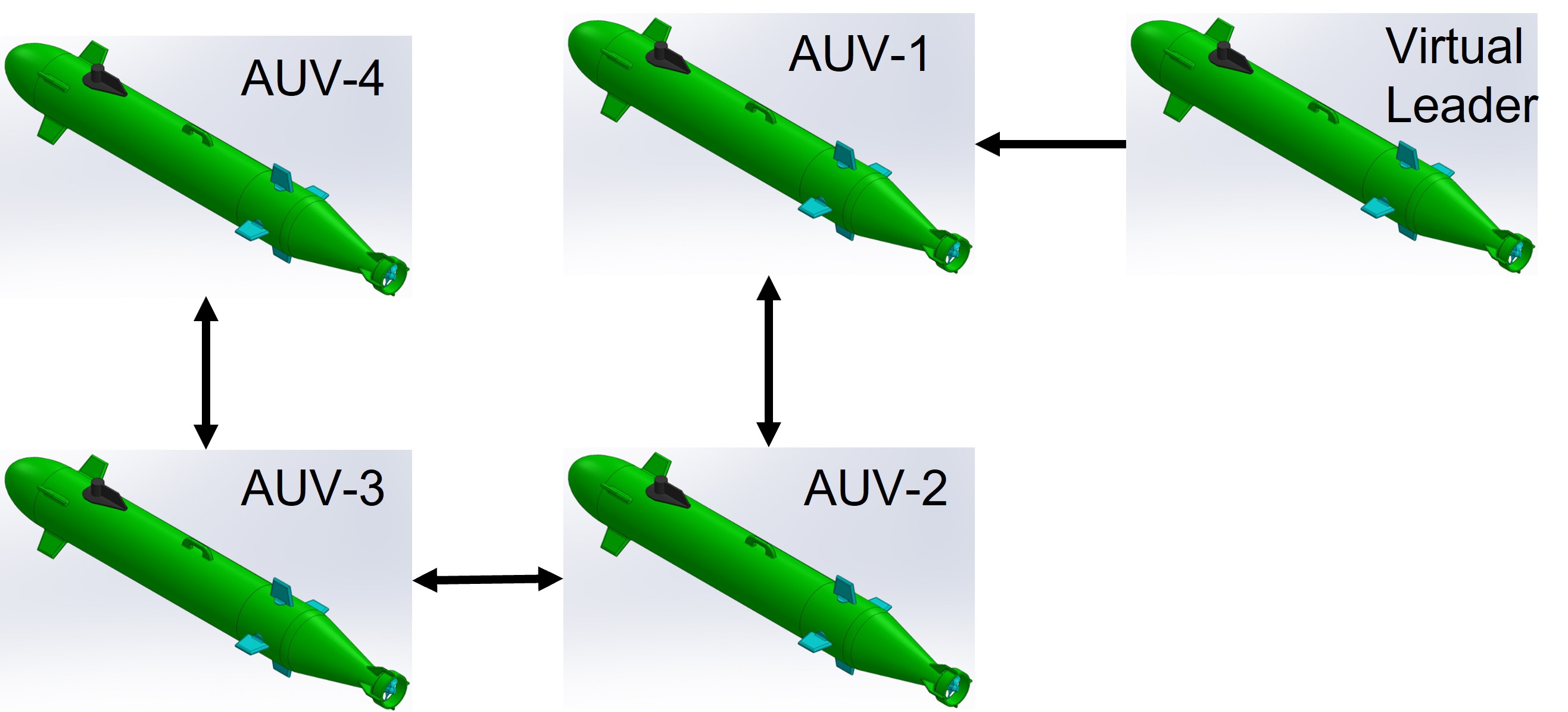}
	\caption{The communication topology graph for 3 AUVs formation control}
\label{fig.1}
\end{figure}

\begin{figure}[!t]\centering
	\includegraphics[width=8.5cm]{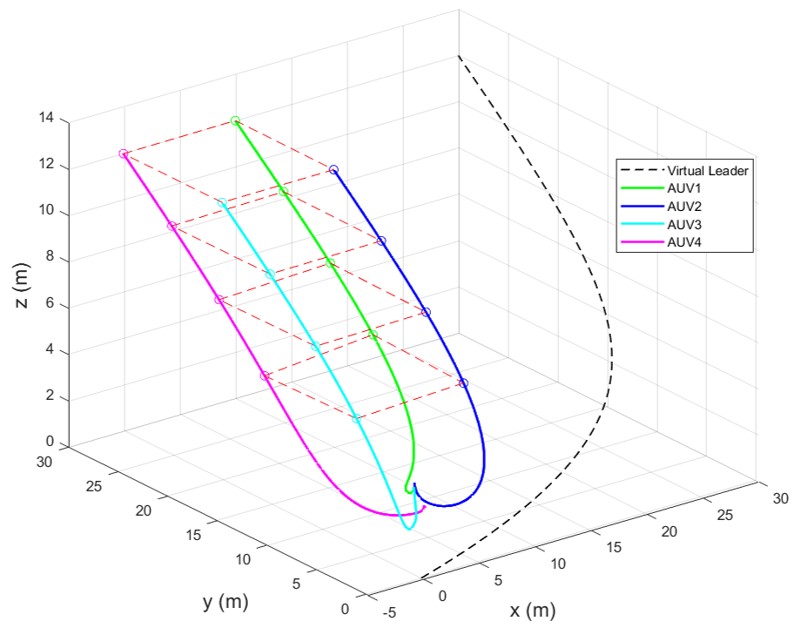}
	\caption{The formation shape of four AUVs under the proposed controller}
\label{fig.2}
\end{figure}

\begin{figure}[!t]\centering
	\includegraphics[width=8.5cm]{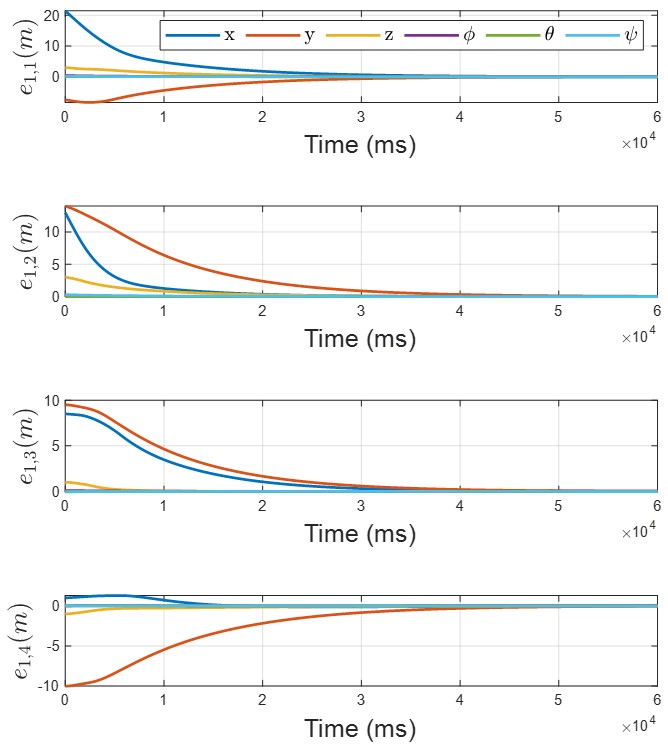}
	\caption{Position tracking error $\varepsilon_{1}={e}_{1}$ of the AUVs under the proposed controller}
\label{fig.3}
\end{figure}

\begin{figure}[!t]\centering
	\includegraphics[width=8.5cm]{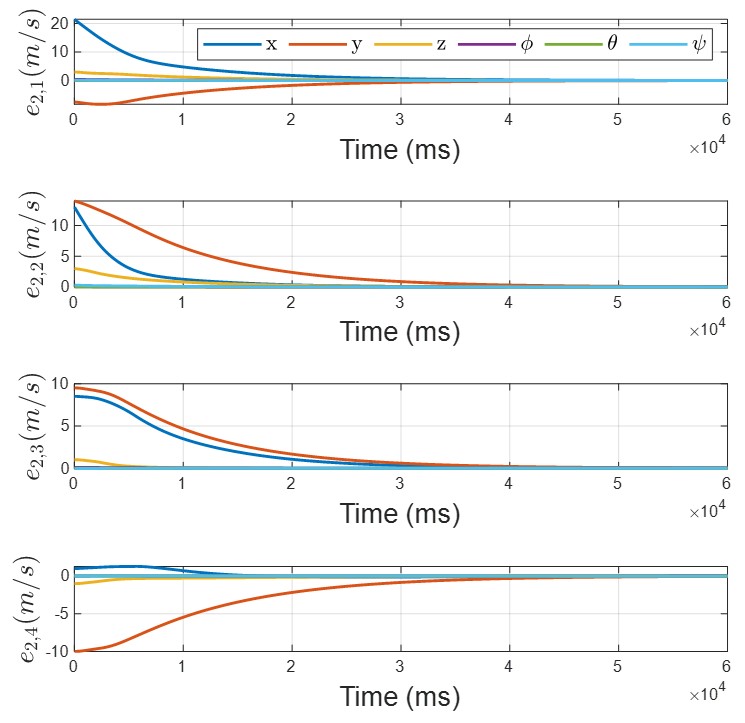}
	\caption{Velocity tracking error $\varepsilon_{2}={e}_{2}$ of the AUVs under the proposed controller}
\label{fig.4}
\end{figure}

\begin{figure}[!t]\centering
	\includegraphics[width=8.5cm]{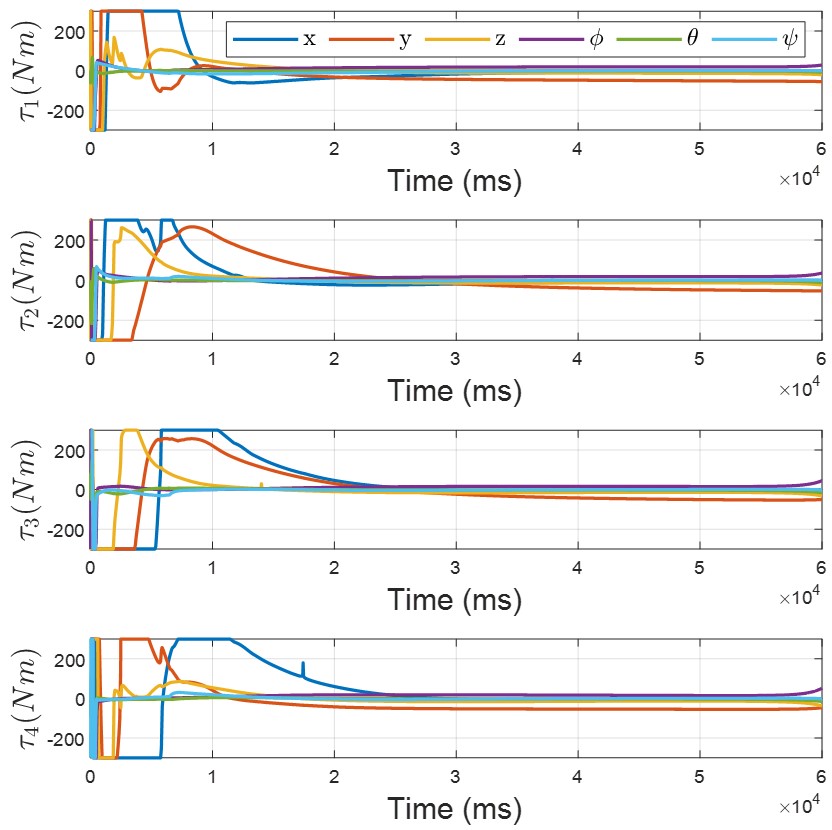}
	\caption{Control efforts $\tau_i (i=1,2,3,4)$ of the AUVs under the proposed controller}
\label{fig.5}
\end{figure}

\begin{figure}[!t]\centering
	\includegraphics[width=8.5cm]{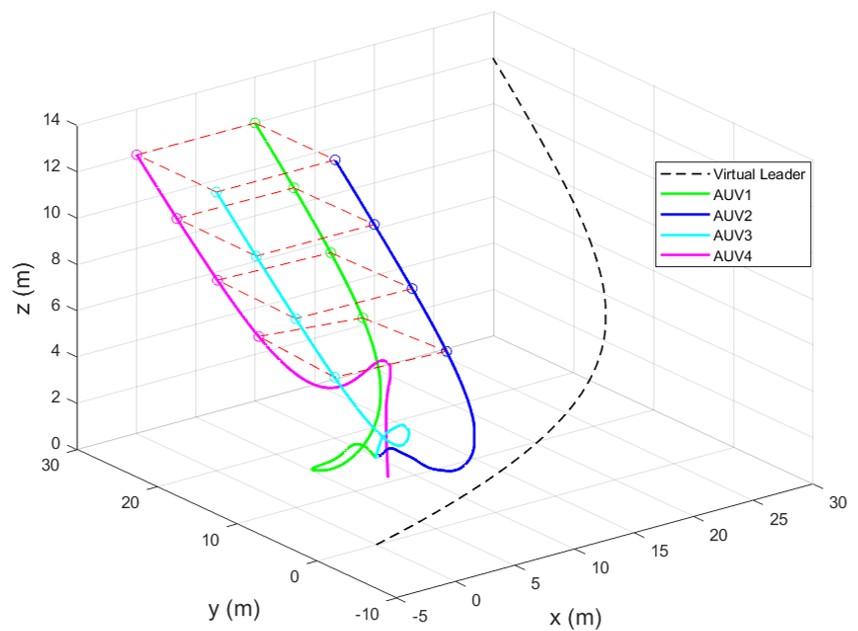}
	\caption{The formation shape of four AUVs under the SMC controller}
\label{fig.6}
\end{figure}

\begin{figure}[!t]\centering
	\includegraphics[width=8.5cm]{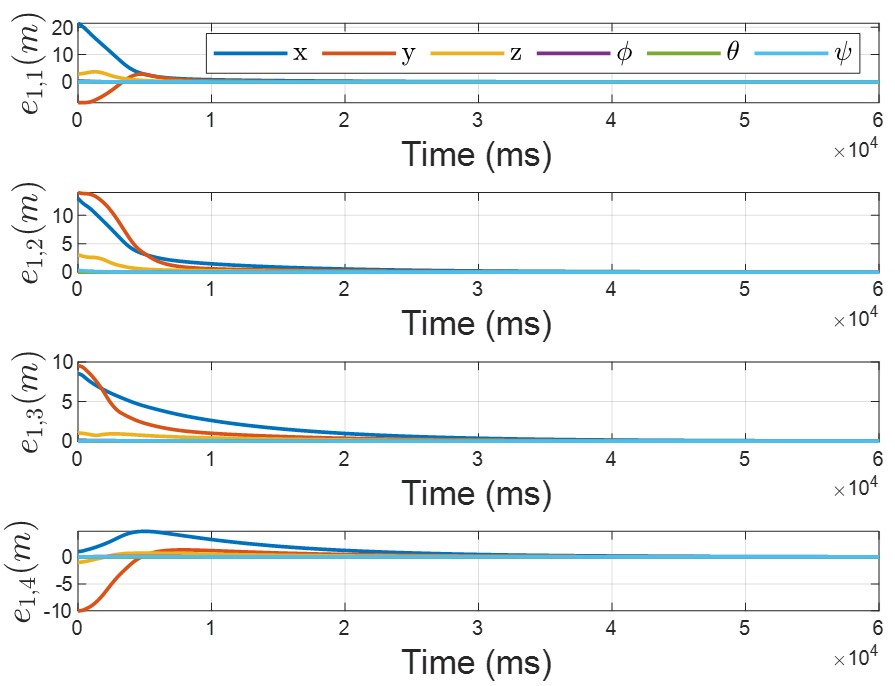}
	\caption{Position tracking errors $\varepsilon_{1}=e_{1}$ of the AUVs under the SMC controller}
\label{fig.7}
\end{figure}

\begin{figure}[!t]\centering
	\includegraphics[width=8.5cm]{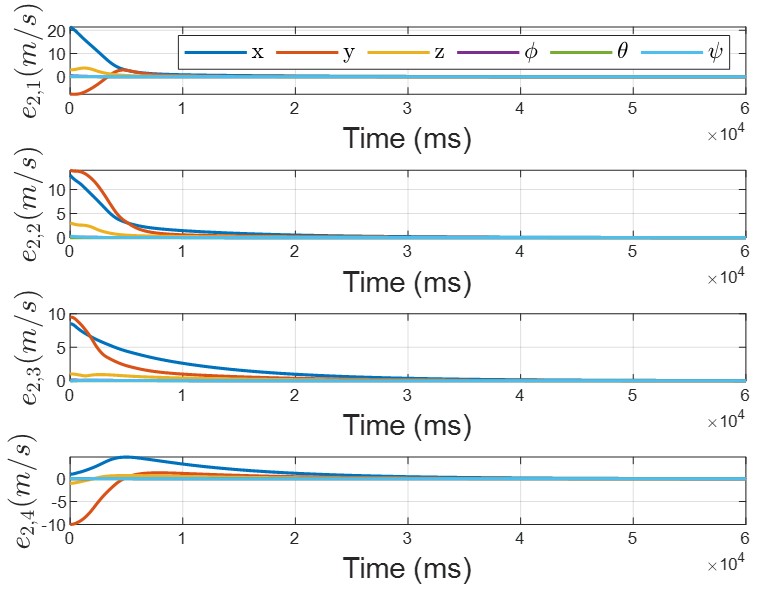}
	\caption{Velocity tracking errors $\varepsilon_{2}=e_{2}$ of the AUVs under the SMC controller}
\label{fig.8}
\end{figure}

\begin{figure}[!t]\centering
	\includegraphics[width=8.5cm]{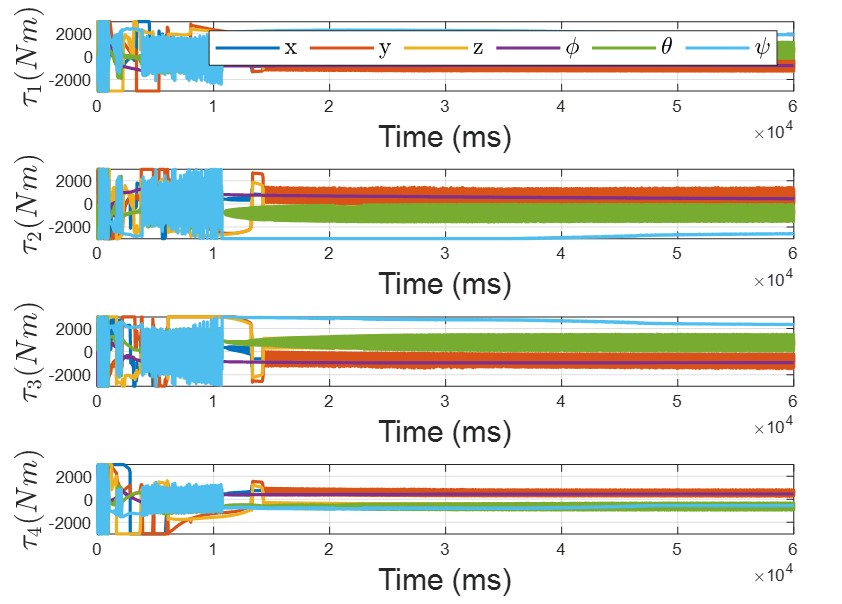}
	\caption{The control efforts $\tau_{i}, (i=1, 2, 3, 4)$ of the AUVs under the SMC controller}
\label{fig.9}
\end{figure}

\section{Conclusion}
This paper has introduced a new distributed fixed-time consensus formation control for multiple AUV systems based on an adaptive backstepping fuzzy sliding mode control. In this scheme, a distributed consensus tracking error was derived to form a distributed global error dynamics. Then, a fixed-time backstepping SMC was designed to obtain a fixed-time convergence for the whole system. However, the backstepping fixed-time SMC had two main shortcomings: (i) it produces a greater magnitude of chattering due to the selection of bigger sliding gain, (ii) it has not considered the input saturation effects in the design. To overcome these shortcomings, an auxiliary variable and an adaptive fixed-time fuzzy logic approximation have been employed. The computer simulation on a formation consensus control for four AUVs demonstrated that the fixed-time controller provided a quick response and stability for the group of AUVs and can handle the input saturation problem well.

In future works, we will address the multiple constraints problem (i.e., working spaces constraints, states constraints and input constraints simultaneously) and obstacle avoidance problems for multiple collaborative AUVs. An physical experiment system based on the hardware systems of BlueROV2 robots are being developed, and the experimental results will be shown in future works.       


\appendices
{
\section*{Appendix A: Design Distributed sliding mode control}
The SMC can be derived as follows \cite{SMC3}:\\
First, the sliding surface is selected as
\begin{equation}
\begin{aligned}
\label {70}
s=k_1(L+B)\bar{\varepsilon}_1+\bar{\varepsilon}_2
\end {aligned}
\end{equation}
where $k_1$ is a positive constant. 

The sliding mode control can be designed as
\begin{equation}
\begin{aligned}
\label {71}
\tau&=[\bar{J}(\eta_2)\bar{\Pi}]^{-1}\left(-\bar{\Phi}(\upsilon,\eta)\upsilon+\tau^{'}\right)
\end {aligned}
\end{equation}

where $\tau'$ is designed as
\begin{equation}
\begin{aligned}
\label {711}
\tau&=[\bar{J}(\eta_2)\bar{\Pi}]^{-1}\left(-k_1\bar{\varepsilon}_2+\textbf{1}_N\otimes\ddot{\eta}^d-\beta_0\text{sign}(s)\right)
\end {aligned}
\end{equation}

To reduce the chattering, the controller (\ref{711}) is revised as:
\begin{equation}
\begin{aligned}
\label {712}
\tau&=[\bar{J}(\eta_2)\bar{\Pi}]^{-1}\left(-k_1\bar{\varepsilon}_2+\textbf{1}_N\otimes\ddot{\eta}^d-\beta_0\frac{s}{||s||+\epsilon_1}\right).
\end {aligned}
\end{equation}
The convergence and stability of the SMC can be referred to \cite{SMC3}.
}

\end{document}